\documentclass[10pt,a4paper,twoside]{article}
\usepackage{baltlat6}
\usepackage{epsfig}
\usepackage{lscape}
\usepackage{here}
\begin{document}
\ \ \vspace{-0.5mm}

\setcounter{page}{337}
\vspace{-2mm}

\titlehead{Baltic Astronomy, vol.\,17, 337--349, 2008}

\titleb{ACCURACY OF STAR CLUSTER PARAMETERS FROM \\ INTEGRATED
UBVRIJHK PHOTOMETRY}

\begin{authorl}
\authorb{A.~Brid\v{z}ius}{1},
\authorb{D.~Narbutis}{1,2},
\authorb{R.~Stonkut\.{e}}{1},
\authorb{V.~Deveikis}{2} and
\authorb{V.~Vansevi\v{c}ius}{1,2}
\end{authorl}

\begin{addressl}
\addressb{1}{Institute of Physics, Savanori\c{u} 231, Vilnius
LT-02300, Lithuania \\ wladas@astro.lt}
\addressb{2}{Vilnius University Observatory, \v{C}iurlionio 29,
Vilnius LT-03100, Lithuania}
\end{addressl}

\submitb{Received 2008 November 4; revised December 21; accepted
December 23}

\begin{summary} We investigate the capability of the {\it UBVRIJHK}
photometric system to quantify star clusters in terms of age, metallicity
and color excess by their integrated photometry in the framework
of P\'{E}GASE single stellar population (SSP) models. The age-metallicity-extinction
degeneracy was analyzed for various parameter combinations, assuming
different levels of photometric accuracy. We conclude, that most
of the parameter degeneracies, typical to the {\it UBVRI} photometric
system, are broken in the case when the photometry data are supplemented
with at least one infrared magnitude of the {\it JHK} passbands,
with an accuracy better than $\sim$\,0.05~mag. The presented analysis
with no preassumptions on the distribution of photometric errors
of star cluster models, provides estimate of the intrinsic capability
of any photometric system to determine star cluster parameters from
integrated photometry.
\end{summary}


\resthead{Accuracy of star cluster parameters}{A.~Brid\v{z}ius, D.
Narbutis, R. Stonkut\.e et al.}

\sectionb{1}{INTRODUCTION}

Broad-band photometric data of extragalactic star clusters are applicable
to derive their evolutionary parameters via comparison with single
stellar population (SSP) models. However, this procedure is restricted
by the fact that in some parameter domains strong age-metallicity-extinction
degeneracies (see, e.g., Worthey 1994) remain unsolved. For a study
of broad-band color indices, most sensitive to various star and cluster
parameters, see, e.g., Jordi et al. (2006) and Li et al. (2007).

It is known (Anders et al. 2004; Kaviraj et al. 2007) that infrared
and ultraviolet passbands are helpful for breaking the age-metallicity-extinction
degeneracies. In this respect infrared observations have the priority:
in some atmospheric windows they are accessible for ground-based
telescopes, while ultraviolet photometry shorter than 300 nm is
possible only from space. In some extragalactic studies of star
clusters published to date various combinations of optical and/or
near-infrared passbands have been used (e.g., Kodaira et al. 2004;
Hempel \& Kissler-Patig 2004; Fan et al. 2006; Hempel et al. 2007;
Narbutis et al. 2008; Pessev et al. 2008).

However, in comparison to photometry in the optical range, ground-based
infrared photometry usually has larger errors due to variable humidity,
and this requires a careful calibration (e.g., Kodaira et al. 1999;
Kidger et al. 2006). The reliability of secondary standards used
for photometric calibration of wide field images, should be verified
by several independent sources, if they are available (e.g., Narbutis,
Stonkut\.{e} \& Vansevi\v{c}ius 2006).

In the present paper we continue to investigate a possibility of
determining star cluster parameters (age, metallicity and color
excess) by comparison of their integrated color indices with the
SSP models computed with the P\'{E}GASE (v.~2.0; Fioc \& Rocca-Volmerange
1997) code package. In our previous study (Narbutis et al. 2007b,
hereafter Paper~I) it was found that the {\it UBVRI} system enables
us to estimate cluster parameters over a wide range of their values,
when the overall accuracy of color indices is better than $\sim$\,0.03~mag.
In the following we discuss how the adding of the {\it JHK} passbands
affects the accuracy of cluster parameter (age, metallicity and
color excess) determination. We analyze degeneracies at various
accuracy levels of photometry for the same values of cluster parameters
as in Paper~I.

In the similar study Anders et al. (2004) have used the so-called
AnalySED method for determining cluster parameters with a different
approach to the degeneracy problem. In Section 2 we discuss the main
differences between the Anders et al. and our methods and note, that
the parameter degeneracy analysis presented in this study is based
on minimum assumptions.

\sectionb{2}{THE METHOD}

The analysis method used in this study is similar to that of Paper~I.
The SSP models were computed with the P\'{E}GASE program package,
applying its default options and the universal initial mass function
(UIMF; Kroupa 2001). The integrated color indices in all pass-bands
in respect to the $V$-band of SSP models were reddened by taking
into account the dependence of color-excess ratio (e.g., $E_{U-B}/E_{B-V}$)
on color index $B$--$V$ of SSP model and assumed color excess, $E_{B-V}$,
applying the standard extinction law (Cardelli et al. 1989). The
three-parameter (3-D) SSP model grid of $\sim$\,$5\times10^{5}$ models
was constructed at the following nodes: (i) 76 age, $t$, values from
1\,Myr to 20\,Gyr with a constant step of $\log\kern1pt(t/{\rm Myr})
= 0.05$;\footnote{~For the ages $<$\,12~Myr the step was equal to
1\,Myr. The SSP models for the ages larger than the oldest known
globular clusters are used to avoid marginal effects.} (ii) 31
metallicity\footnote{~[M/H] was computed applying the approximation
${\rm [M/H]}=\log\kern1pt(Z/Z_{\sun})$.}, [M/H], values from $-2.3$
to $+0.7$, with a step of 0.1\,dex; (iii) 201 color excess, $E_{B-V}$,
values from 0.0 to 2.0, with a step of 0.01.

The procedure of determination of cluster parameters ($t$, [M/H],
$E_{B-V}$) was implemented as a C++ code in the data analysis and
the graphing software package `Origin' (OriginLab Corporation). It
is based on a similar technique developed for star quantification
by Vansevi\v{c}ius \& Brid\v{z}ius (1994), i.e., the comparison of
the observed color indices of a star cluster with color indices of
the SSP from the model grid. For this purpose we use the quantification
quality criterion, $\delta$, calculated by the formula:
\begin{equation}
{\delta=\sqrt{\sum_{}^{}w_{i}(CI_{i}^{\rm obs}-CI_{i}^{\rm mod})^{2}\over\sum_{}^{}w_{i}}}\,,
\end{equation}
where $CI_{i}^{\rm obs}$ stands for the ``observed'' color indices
$U$--$V$, $B$--$V$, $V$--$R$, $V$--$I$, $V$--$J$, $V$--$H$ and $V$--$K$;
$CI_{i}^{\rm mod}$ -- the corresponding color indices of the SSP
models from the grid; $w _{i}$ -- the weights for the observed color
indices.

We investigate the possibility to determine cluster parameters using
the {\it UBVRIJHK} photometric system for 54 models taken from the
SSP model grid as ``observed'' objects with the following parameters:
$t=0.02$, 0.05, 0.1, 0.2, 0.5, 1, 2, 5, 10~Gyr, [M/H]~=~0.0, $-0.7$,
$-1.7$ and $E_{B-V}=0.1$, 1.0. Their color indices are denoted by
$CI_{i}^{\rm obs}$, see Eq.~1. We have assigned the weights of $w_{i}
= 1$ for $U-V$, $B-V$, $V-R$ and $V-I$ color indices, and $w_{i} =
1/9$ for $V$--$J$, $V$--$H$ and $V$--$K$ of the ``observed'' star
clusters. Such weighting scheme assumes that the accuracy of the
infrared {\it JHK} observations is $\sim$\,3 times lower than that
of the optical {\it UBVRI} -- a typical condition in photometry of
extragalactic star clusters.

Anders et al. (2004) have extensively analyzed the possibilities of
their AnalySED method, which has a different approach to the degeneracy
problem, comparing to the method presented in this paper. First,
AnalySED is based on the comparison of observed magnitudes with the
model magnitudes, while for the same aim we are using color indices.
This allows us to avoid fitting of an additional free parameter --
the cluster mass. Second, Anders et al. (2004) define the ``observed''
magnitudes of model clusters by assigning to them random errors.
However, we did not apply ``observation errors'' to color indices,
since it is reasonable to assume that the quantification quality
criterion, $\delta$, reflects the effect of photometric accuracy
sufficiently well. In general, $\delta$ represents the lowest possible
average photometric error of all color indices, which in the case
of real clusters can be larger, mostly due to the influence of crowding
by contaminating field stars (e.g., Narbutis et al. 2007a). Therefore,
the parameter degeneracy analysis presented in this study is based
on minimum assumptions.

We have used several $\delta_{\rm max}$ values as the upper threshold
levels of photometric errors. The cluster parameters were determined
independently at each accuracy level by averaging the corresponding
parameter values of SSP models at the grid nodes, which have $\delta
\leq \delta_{\rm max}$. The weights for the calculation of the parameter
averages and the standard deviations were assigned 1 and $10^{-4}/\delta^{2}$
for the nodes with $\delta\leq0.01$ and $>$\,0.01~mag, respectively.

\begin{figure}[!th]
\centerline{\psfig{figure=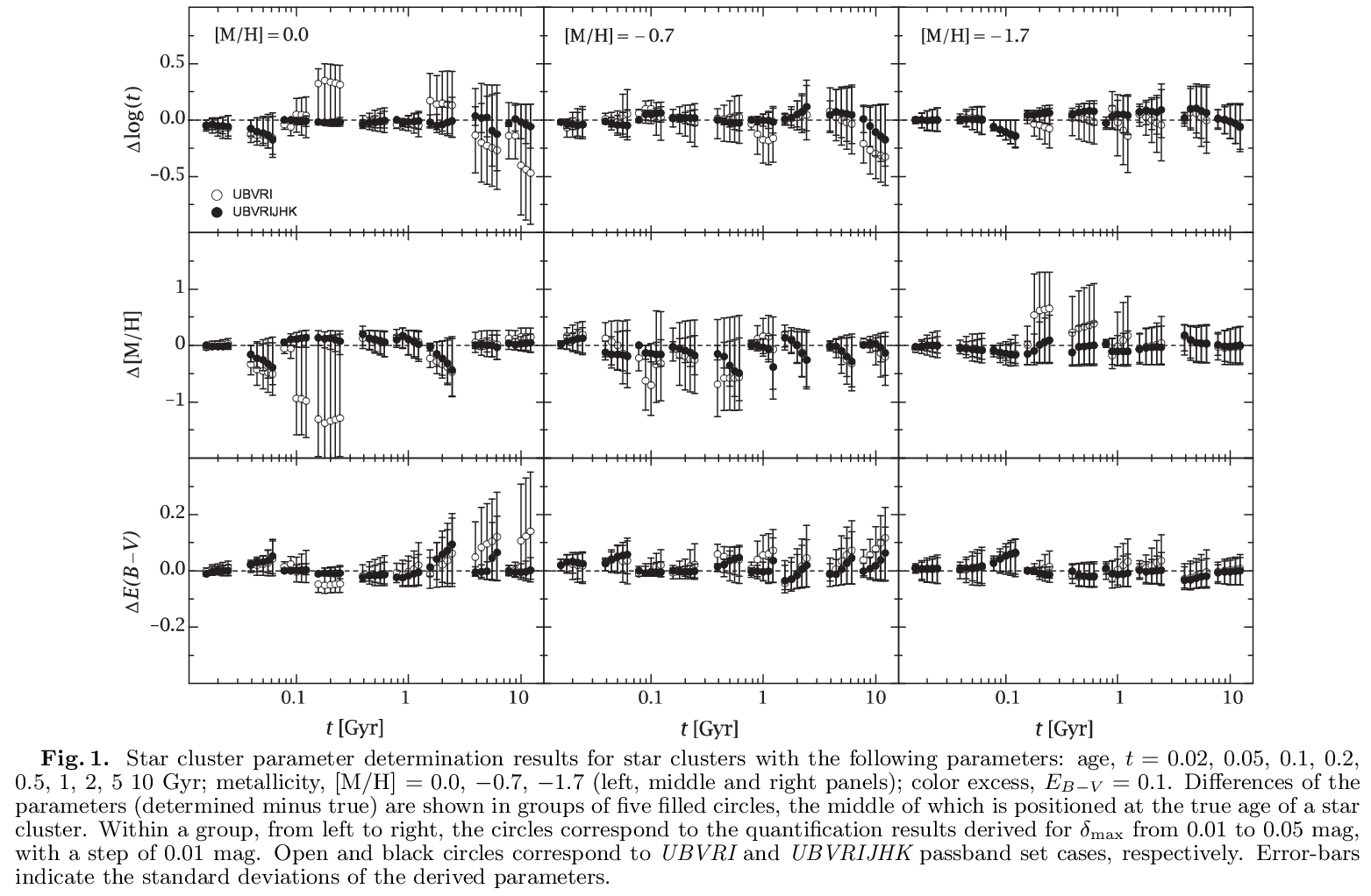,width=185truemm,angle=90,clip=}}
\end{figure}

\begin{figure}[!th]
\centerline{\psfig{figure=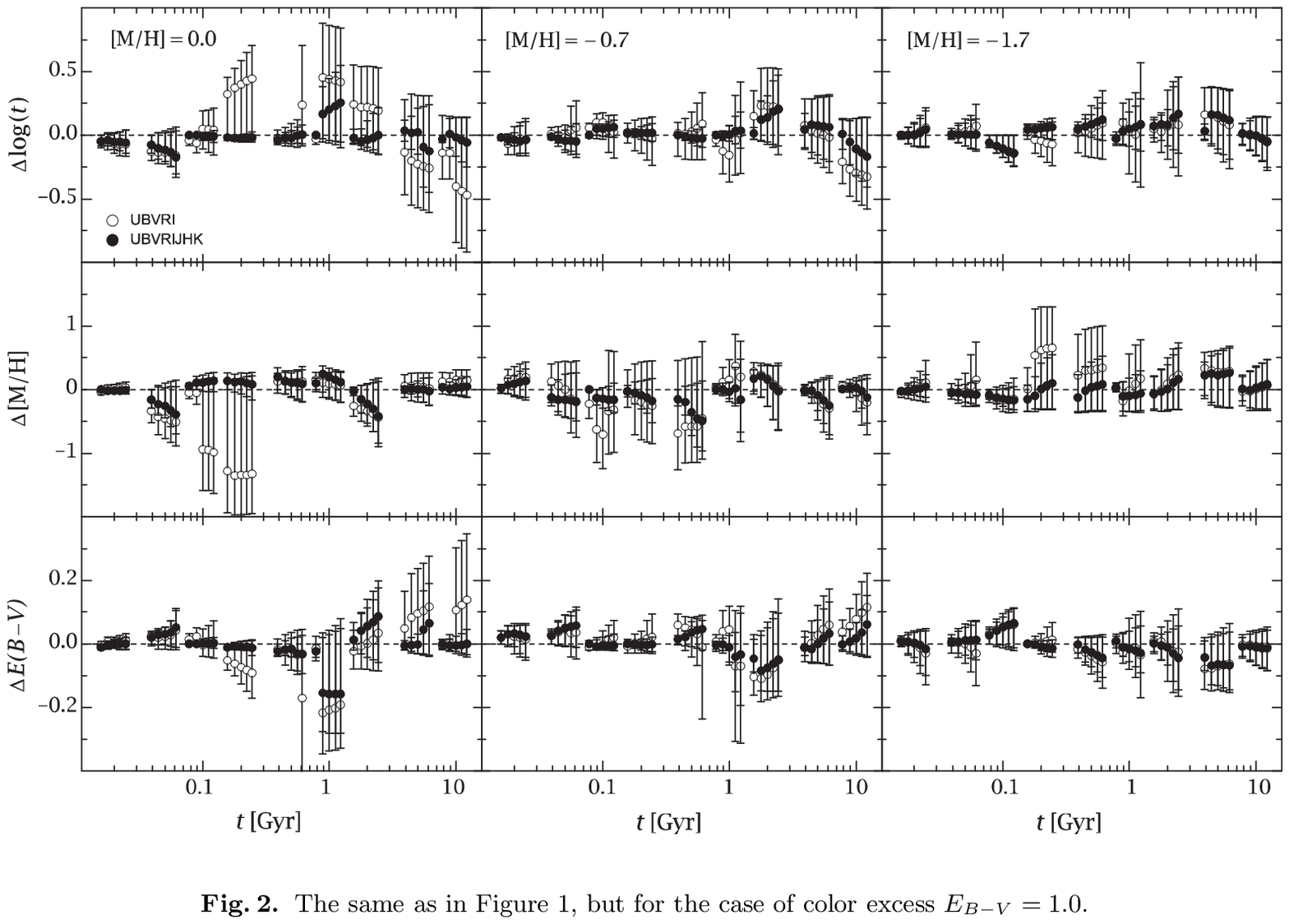,width=185truemm,angle=90,clip=}}
\end{figure}

\begin{figure}[!t]
\vbox{\centerline{\psfig{figure=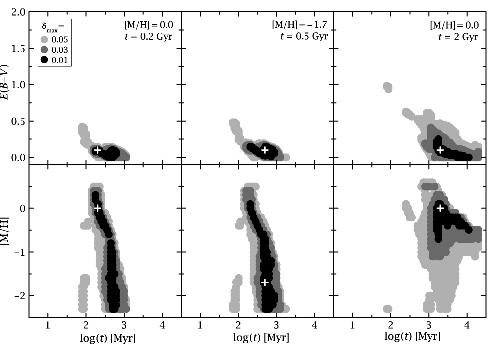,width=120truemm,angle=0,clip=}}
\captionb{3}{The distribution of the quantification quality criterion,
$\delta$, in projection onto the $E_{B-V}$ and [M/H] vs.~$\log\kern1pt(t/{\rm
Myr})$ planes for the clusters with the parameters: $E_{B-V}=0.1$,
[M/H]~=~0.0, $t=0.2$ and 2~Gyr (left and right panels); $E_{B-V}=0.1$,
[M/H]~=~$-1.7$, $t=0.5$~Gyr (middle panel) for the case of {\it UBVRI}
passband set. The true positions of cluster parameters are marked
by a white ``plus'' symbol. Black, gray and light-gray shaded areas
correspond to $\delta_{\rm max}=0.01$, 0.03 and 0.05~mag, respectively.}}
\end{figure}

\begin{figure}[!th]
\vbox{\centerline{\psfig{figure=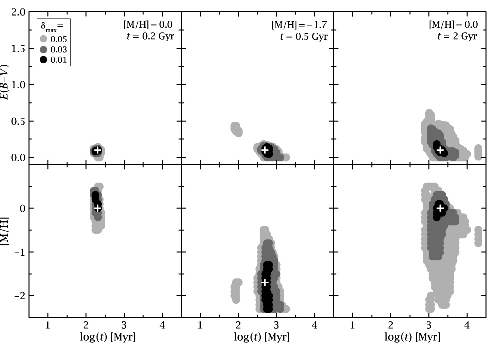,width=120truemm,angle=0,clip=}}
\vspace{-1mm}
\captionb{4}{The same as in Figure~3, but for the case of {\it
UBVRIJHK} passband set.}}
\vspace{3mm}
\end{figure}

\begin{figure}[!th]
\vbox{\centerline{\psfig{figure=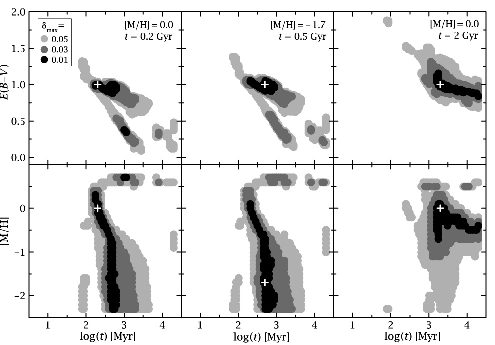,width=120truemm,angle=0,clip=}}
\vspace{-1mm}
\captionb{5}{The same as in Figure~3 for the case of {\it UBVRI}
passband set, but for $E_{B-V}=1.0$.}}
\vspace{3mm}
\end{figure}

\begin{figure}[!th]
\vbox{\centerline{\psfig{figure=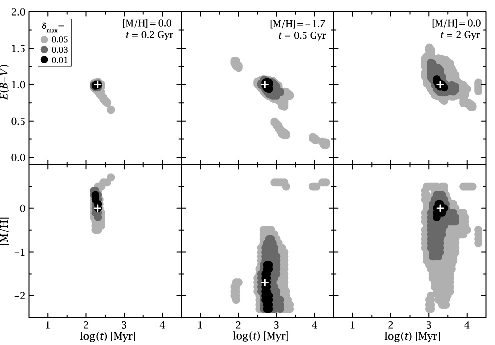,width=120truemm,angle=0,clip=}}
\vspace{-1mm}
\captionc{6}{The same as in Figure~5, but for the case of {\it
UBVRIJHK} passband set.}}
\vspace{3mm}
\end{figure}

\begin{figure}[!th]
\vbox{\centerline{\psfig{figure=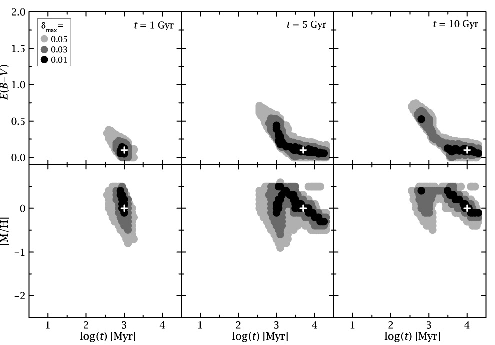,width=120truemm,angle=0,clip=}}
\vspace{-1mm}
\captionb{7}{The same as in Figure~3 for the case of {\it UBVRI}
passband set, but for the star clusters with the following parameters:
$E_{B-V}=0.1$, [M/H]~=~0.0, $t$ = 1, 5 and 10~Gyr.}}
\vspace{3mm}
\end{figure}

\begin{figure}[!th]
\vbox{\centerline{\psfig{figure=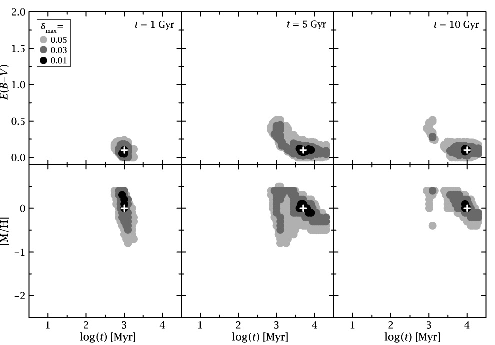,width=120truemm,angle=0,clip=}}
\vspace{-1mm}
\captionc{8}{The same as in Figure~7, but for the case of {\it
UBVRIJHK} passband set.}}
\vspace{4mm}
\end{figure}

\begin{figure}[!th]
\vbox{\centerline{\psfig{figure=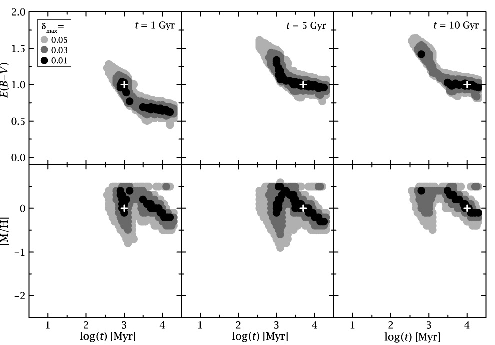,width=120truemm,angle=0,clip=}}
\captionb{9}{The same as in Figure~7 for the case of {\it UBVRI}
passband set, but for $E_{B-V}=1.0$.}}
\end{figure}

\begin{figure}[!th]
\vbox{\centerline{\psfig{figure=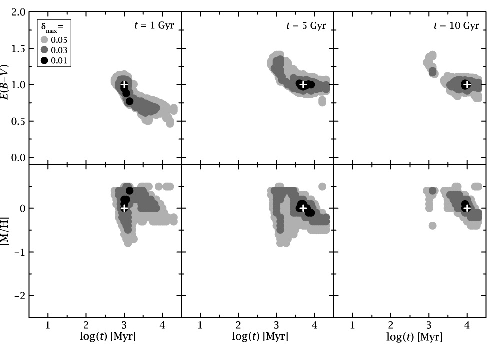,width=120truemm,angle=0,clip=}}
\vspace{-1mm}
\captionc{10}{The same as in Figure~9, but for the case of {\it
UBVRIJHK} passband set.}}
\end{figure}

\begin{figure}[!th]
\centerline{\psfig{figure=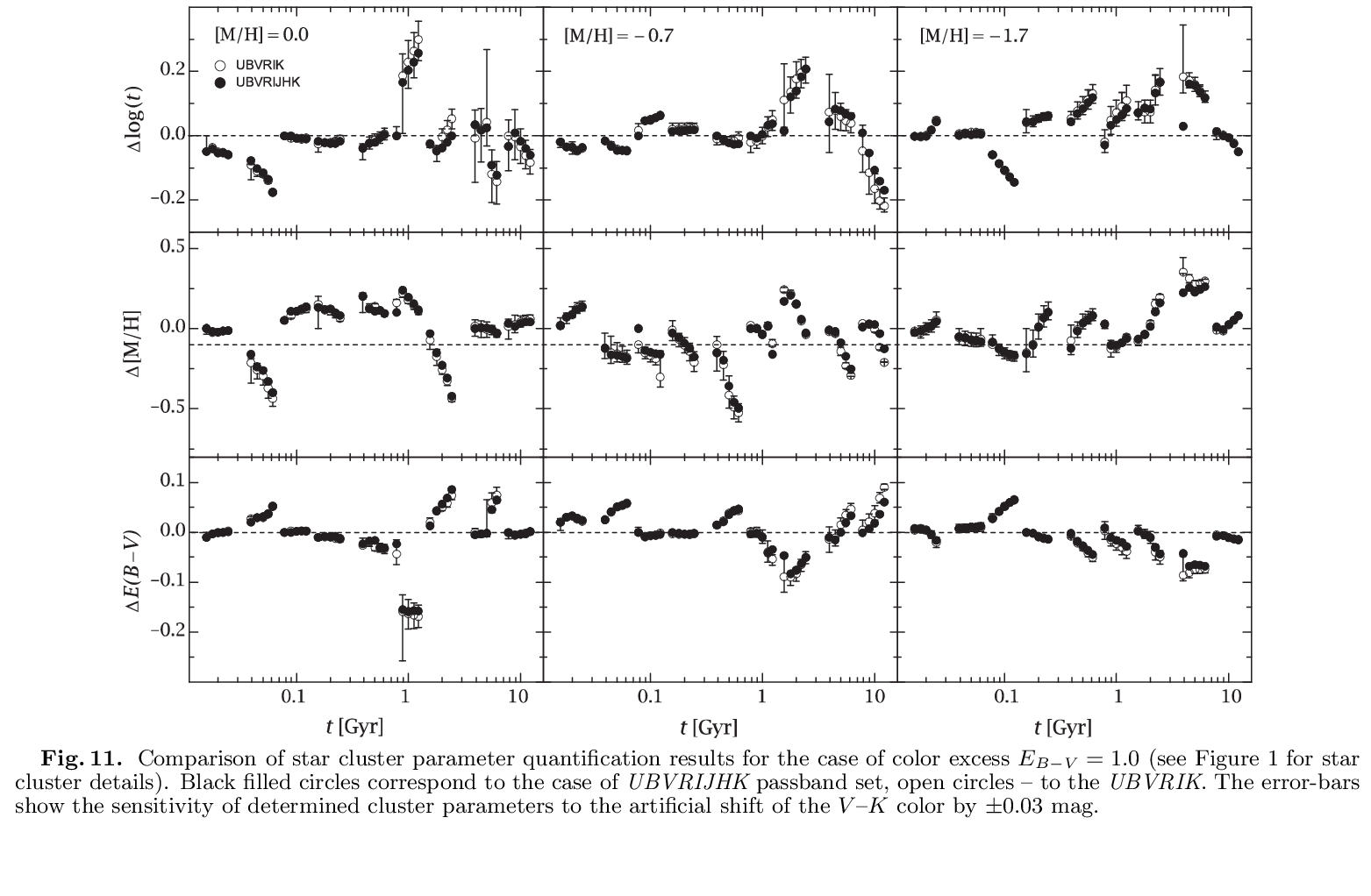,width=185truemm,angle=90,clip=}}
\end{figure}

Furthermore, for the parameter averaging we used only the nodes,
which reside in a single continuous `island' around the true cluster
position in the 3-D parameter space of the SSP model grid. Boundaries
of `islands' at each $\delta_{\rm max}$ level were determined automatically
by the {\it clustering procedure}, which finds discontinuities in
the parameter space, starting from the true position of the cluster.
Such a procedure excludes nodes, which are located in the secondary
$\delta$ minima, arising due to the age-metallicity-extinction
degeneracies in the 3-D parameter space.

\sectionb{3}{RESULTS AND DISCUSSION}

The results of parameter determinations of clusters are provided
in Figures~1 and 2 for color excess values $E_{B-V}=0.1$ and 1.0,
respectively. In each panel the differences of parameters (determined
minus true) are shown in groups of five filled circles, the middle
circle is positioned at the true age of the cluster, i.e., the circles
indicate parameters determined at $\delta_{\rm max}$ threshold values
from 0.01 to 0.05~mag, with a step of 0.01~mag, plotted from left
to right. Open and black circles correspond to the {\it UBVRI} and
{\it UBVRIJHK} passband set cases, respectively. Error-bars indicate
 standard deviations of the determined parameters and characterize
the integrated `size' of the $\delta\leq\delta_{\rm max}$ `island'
in the corresponding parameter space.

It is clearly visible in both Figures~1 and 2, that additional usage
of {\it JHK} passbands significantly improves the precision of cluster
parameter determination with respect to the {\it UBVRI} passbands
alone. Especially good results are in the case of the small color
excess, $E_{B-V}=0.1$, value. Note, that strong age-metallicity
degeneracies at ages $t=0.1$, 0.2 and 10~Gyr, which are present in
the {\it UBVRI} case, completely disappear in the {\it UBVRIJHK}
case. A similar, but a slightly weaker effect is also visible for
models of ages $t=2$ and 5~Gyr. However, if objects are highly reddened,
$E_{B-V}=1.0$, there are still notable parameter degeneracies for
some cases. The strongest age-extinction degeneracy is for clusters
with: $t=1$~Gyr and [M/H]~=~0.0; $t=2$~Gyr and [M/H]~=~$-0.7$; $t=5$~Gyr
and [M/H]~=~$-1.7$. Actually, the quantification results of the last
case are much more accurate, when SSP models older than 15~Gyr are
not included in the calculation of the average.

The impact of adding the {\it JHK} passbands to the {\it UBVRI} system
on the accuracy of the quantification of cluster parameters is illustrated
in Figures 3--10 for the clusters of ages $t=0.2$, 1, 2, 5, 10~Gyr
for solar metallicity, and $t=0.5$~Gyr for [M/H]~=~$-1.7$. The figures
display cluster parameter quantification maps at the threshold levels
of $\delta_{\rm max}=0.01$, 0.03 and 0.05~mag.

In Figures~3 and 5 for the {\it UBVRI} set, the left and the central
panels look very similar, although the left panel displays quantification
map of a cluster with $t=200$~Myr and [M/H]~=~0.0, while the central
panel -- for a cluster with $t=500$~Myr and [M/H]~=~$-1.7$. This
means that due to age-metallicity degeneracy for both clusters we
obtain similar and wrong parameters. However, the situation changes,
when the {\it JHK} magnitudes are added. Figures~4 and 6 display
quantification maps of the same clusters, but for the {\it UBVRIJHK}
passband set. Note, that degeneracies disappear, even at $\delta_{\rm
max}=0.05$~mag level.

An interesting example is the cluster with the parameters $t=2$~Gyr
and [M/H]~=~0.0. Figures~1 and 2 show that the addition of the {\it
JHK} magnitudes improves the age determination, but $E_{B-V}$ determination
becomes of lower accuracy, comparing to the {\it UBVRI} set alone.
Rightmost panels in Figures~3--6 clarify this situation. In the {\it
UBVRI} case, strong age-metallicity degeneracy takes place: note the
additional `tail' of suitable SSP models in the $E_{B-V}$ vs.~$t$
plane towards older ages and lower $E_{B-V}$ values. This helps to
compensate the influence of SSP models at higher $E_{B-V}$ values,
and the quantification produces a `correct' reddening. In the case
of {\it UBVRIJHK}, no such `tail' is seen, therefore the determined
$E_{B-V}$ value is shifted from its true position.

Figures~7--10 display three similar cases of parameter degeneracies
for clusters of solar metallicity and ages $t=1$, 5 and 10~Gyr. Only
the 10 Gyr age sample is sensitive enough to the addition of the
{\it JHK} passbands, when parameter degeneracies become broken even
at $\delta_{\rm max}=0.05$~mag level. On the contrary, for the ages
of 1~Gyr (except for the case of $E_{B-V}=0.1$; see Figures~7 and
8) and 5~Gyr, parameter degeneracies exist even for $\delta_{\rm
max}=0.01$ and 0.03~mag, respectively. This implies, that the overall
accuracy of cluster color indices must be better than 0.03~mag, which
is rather difficult to achieve in infrared photometry.

We also investigated separately the influence of each infrared passband
to the parameter determination precision with the purpose to find
a minimum passband set, which would be sufficient for the quantification
of clusters. The result is that all three color indices ($V$--$J$,
$V$--$H$, $V$--$K$), added to the {\it UBVRI} photometric system
separately, reduce the quantification errors, but $V$--$K$ is most
important. Figure~11 displays the quantification results for clusters
using the {\it UBVRIJHK} (filled circles) and the {\it UBVRIK} (open
circles) passband sets, both pictures are quite similar. Thus the
{\it K} passband helps to solve majority of the parameter degeneracy
problems. The error-bars in the {\it UBVRIK} case show the sensitivity
of the parameters to the change of $V$--$K$ by $\pm$\,0.03~mag.

Recently an updated version of stellar isochrones, compared to those
used in P\'{E}GASE, has been released by the Padova group, which
includes a more accurate treatment of thermally pulsating AGB stars
(Marigo et al. 2008). It has been shown that the interpretation of
integrated colors of unresolved galaxies, by applying galaxy populations
synthesis method and the new isochrone set, can be significantly
altered due to improved models of AGB stars (e.g., Tonini et al.
2008). In the case of single stellar population analysis of star
clusters one can expect similar changes. Our method of the parameter
degeneracy analysis, which is much simpler than the methods described
in the literature (see, e.g., de Grijs et al. 2005), can be used
to study the effects AGB stars on cluster colors by intercomparing
parameter degeneracies in various SSP model frameworks.

\sectionb{4}{CONCLUSIONS}

We conclude that additional photometric information from the {\it
JHK} passbands can significantly improve the accuracy of the determination
of cluster parameters based on the P\'{E}GASE SSP models, in comparison
with the results when photometry only the {\it UBVRI} passbands is
available. Even one additional {\it K} passband can improve significantly
the capability of the {\it UBVRI} photometric system to eliminate
age-metallicity and age-extinction degeneracies in the majority of
the investigated cluster models, when the overall accuracy of color
indices is better than $\sim$0.05~mag.

Note, however, that this condition of photometric accuracy is broken
for young low-mass star clusters, where the stochastic effects,
arising due to a few bright stars, dominate (e.g., Cervi\~{n}o \&
Luridiana 2004; Deveikis et al. 2008). This implies, that even in
case of ideal photometric calibrations, the uncertainty of cluster
colors due to stochastic effects limits the applicability of SSP
model fitting to derive evolutionary parameters of low-mass star
clusters.

\vskip3mm
\enlargethispage{5mm}

\thanks{We thank the anonymous referee for critical comments and
numerous suggestions, which helped to improve the paper. This work
was financially supported in part by a Grant of the Lithuanian State
Science and Studies Foundation.}

\References

\refb Anders~P., Bissantz~N., Fritze-v.~Alvensleben~U., de~Grijs~R.
2004, MNRAS, 347, 196

\refb Cardelli~J.~A., Clayton~G.~C., Mathis~J.~S. 1989, ApJ, 345,
245

\refb Cervi\~{n}o~M., Luridiana~V. 2004, A\&A, 413, 145

\refb Deveikis~V., Narbutis~D., Stonkut\.{e}~R., Brid\v{z}ius~A.,
Vansevi\v{c}ius~V. 2008, Baltic Astronomy 17, 351

\refb Fan~Z., Ma~J., de Grijs~R., Yang~Y., Zhou~X. 2006, MNRAS, 371,
1648

\refb Fioc~M., Rocca-Volmerange~B. 1997, A\&A, 326, 950

\refb de Grijs~R., Anders~P., Lamers~H.\,J.\,G.\,L.\,M., Bastian~N.
et al. 2005, MNRAS, 359, 874

\refb Hempel~M., Kissler-Patig~M. 2004, A\&A, 428, 459

\refb Hempel~M., Zepf~S., Kundu~A., Geisler~D., Maccarone~T.~J. 2007,
ApJ, 661, 768

\refb Jordi~C., H{\o}g E., Brown A.\,G.\,A., Lindegren L. et al.
2006, MNRAS, 367, 290

\refb Kaviraj~S., Rey~S.-C., Rich~R.~M., Yoon~S.-J., Yi~S.~K. 2007,
MNRAS, 381, L74

\refb Kidger~M.~R., Martin-Luis~F., Artigue~F., Gonzalez-Perez~J.~N.,
Perez-Garcia~A., Narbutis~D. 2006, The Observatory, 126, 166

\refb Kodaira~K., Vansevi\v{c}ius~V., Tamura~M., Miyazaki~S. 1999,
ApJ, 519, 153

\refb Kodaira~K., Vansevi\v{c}ius~V., Brid\v{z}ius~A., Komiyama~Y.,
Miyazaki~S., Stonkut\.{e}~R., \v{S}ablevi\v{c}i\={u}t\.{e}~I.,
Narbutis~D. 2004, PASJ, 56, 1025

\refb Kroupa~P. 2001, MNRAS, 322, 231

\refb Li~Z., Han~Z., Zhang~F. 2007, A\&A, 464, 853

\refb Marigo~P., Girardi L., Bressan A. et al. 2008, A\&A, 482, 883

\refb Narbutis~D., Stonkut\.{e}~R., Vansevi\v{c}ius~V. 2006, Baltic
Astronomy, 15, 471

\refb Narbutis~D., Vansevi\v{c}ius~V., Kodaira~K., Brid\v{z}ius~A.,
Stonkut\.{e} R. 2007a, Baltic Astronomy, 16, 409

\refb Narbutis~D., Brid\v{z}ius~A., Stonkut\.{e}~R., Vansevi\v{c}ius~V.
2007b, Baltic Astronomy, 16, 421 (Paper~I)

\refb Narbutis~D., Vansevi\v{c}ius~V., Kodaira~K., Brid\v{z}ius~A.,
Stonkut\.{e}~R. 2008, ApJS, 177, 174

\refb Pessev~P.~M., Goudfrooij~P., Puzia~T.~H., Chandar~R. 2008,
MNRAS, 385, 1535

\refb Tonini~C., Maraston~C., Devriendt~J., Thomas~D., Silk~J. 2008,
arXiv:0812.1225

\refb Vansevi\v{c}ius~V., Brid\v{z}ius~A. 1994, Baltic Astronomy,
3, 193

\refb Worthey~G. 1994, ApJS, 95, 107

\end{document}